# Fluctuations of Transmission Distribution in Disordered Conductors

Yuli V. Nazarov

*Faculteit der Technische Natuurkunde, Technische Universiteit Delft, 2628 CJ Delft, the Netherlands*

We developed a microscopic approach to calculate the sample-to-sample fluctuations of transmission distribution in disordered conductors. This bridges between Green's function and random matrix theories of quantum transport. The results obtained show that the correlations of transmission eigenvalues obey universal Dyson statistics at small separations between eigenvalues being non-universal otherwise. The results facilitate an easy computation of different physical quantities and impose important constrains on a future imaginable theory of quantum transport.

The concept of universal conductance fluctuations [1] proves to be one of the most interesting concepts in mesoscopic physics. Quantum interference results in conductance of a disordered coherent conductor fluctuates from sample to sample by a value of the order of $G_Q = e^2/(2\pi\hbar)$. The exhaustive qualitative description of the phenomenon has been elaborated on the basis of Green's function technique. [2]

The alternative explanation of UCF has been proposed by Imry [3], who asserted universal explanation of the phenomenon. The conductance of a mesoscopic sample can be expressed as a sum of the eigenvalues of the transmission matrix square, $\mathbf{T} \equiv \mathbf{t}^\dagger \mathbf{t}$,

$$G = \frac{e^2}{\pi\hbar}\mathrm{Tr}(\mathbf{t}^\dagger \mathbf{t}) = \frac{e^2}{\pi\hbar}\sum_n T_n \quad (1)$$

Basing on works by Pichard, [4] Imry showed that the assumption of Wigner-Dyson statistics of transmission eigenvalues yields the correct order of magnitude of UCF.

The pioneering paper by Imry has triggered persistent attempts to formulate the theory of quantum transport as a kind of random matrix theory (RMT). This theory would produce the quantitative description of transmission matrix being based on a few general postulates. [5] However such a theory was doomed to fail. Green's function methods clearly showed that UCF are not precisely universal. Their magnitude depends on the shape of the sample and its dimensionality. Universal RMT would never have such input parameters.

The puzzle is that the RMT can be made working. It may correctly describe quasi-one-dimensional geometry [6,7] and chaotic cavities. [8] Why? How could it be that the same phenomenon combines universal and non-universal features? Are there ways to an improved RMT which would able to describe a concrete physical situation?

Below we answer these questions checking the Imry conjecture by the recently formulated [9] Green's function method. The short answer can be formulated as follows: correlations between transmission eigenvalues obey Wigner-Dyson statistics provided their separation is much smaller than unity. At separations of the order of unity their correlations become dependent on a concrete geometry of a sample. When calculating the fluctuation of conductance, one sums up over small and moderate separations. This is why the final answer combines universal and sample-dependent features. The effective interaction between transmissions that corresponds to the correlations calculated, would be an input parameter of the imaginable improved RMT.

One can draw a parallel between statistics of transmission eigenvalues through an open systems and statistics of energy levels in a closed system. The latter was shown [10,11] to be universal at small energy separation and sample-dependent at energy differences $\equiv E_{th}$.

To proceed, we make use of multicomponet Green's function method proposed in [9]. It expresses generating function of momenta of the transmission distribution

$$F(\Phi) = \mathrm{Tr}\Big[\frac{\mathbf{T}}{1 - \sin^2(\Phi/2)\mathbf{T}}\Big) \quad (2)$$

in terms of an artificially constructed Green's function which is a 2×2 matrix. energy. The Green's function introduced can be readily averaged over impurity configurations with using the conventional technique. [12]

To find the correlations between transmissions, we have to compute the average product of two Green's functions of the kind. The technical details will be reported elsewhere. [13] The corresponding diagrams preserve the common cooperon diffusion structure [1]. However, those are difficult to handle due to the extra matrix structure and coordinate dependence of the average Green's function. The compact answer is possible to obtain for the expressions of the form (2). Using a common diagrammatic trick outlined in [14], we obtain

$$\sin\Phi\sin\Phi' <F(\Phi)F(\Phi')> = 4\frac{\partial}{\partial\Phi}\frac{\partial}{\partial\Phi'}(\ln\det(\check{1}-\check{K}^d)+\ln\det(\check{1}-\check{K}^c)). \quad (3)$$

The operator $\check{K}^{d(c)}$ is actually an integral kernel corresponding to a diffusion (cooperon) ladder section. For a common case of contact impurity potential characterized by the elastic scattering time $\tau$ it reads

$$K^d_{ab,cd}(x,x';\Phi,\Phi') = \frac{1}{\tau(x)}<G^{ac}(x,x';\Phi)><G^{ad}(x,x';\Phi')>, \quad (4)$$

$$K^c_{ab,cd}(x,x';\Phi,\Phi') = \frac{1}{\tau(x)}<G^{ac}(x,x';\Phi)><G^{da}(x',x;\Phi')>, \quad (5)$$

with indexes labeling spin and the additional matrix structure

To solve the problem, we have to find the eigenvalues of the operators $\check{1}-\check{K}^{d(c)}, \mu$. It is possible to derive diffusion-like equations for those. These equations can be solved for a slab geometry of the conductor. In this case $\mu$ can be readily expressed in terms of common diffusion and cooperon mode eigenvalues.

These modes have been intensively studied, [12], and I shortly remind the reader the results. Let the slab dimensions be $L_x, L_y, L_z$, $x$ being transport direction. First, the discrete diffusion and cooperon modes can be labeled by three integer numbers $n_{x,y,z}$, $n_{y,z}$ ranging from 0 to $\infty$ and $n_x$ starting from 1. The eigenvalues normalized by Thouless energy $D/L_x^2$ are given by ( $\vec{n}=(n_x,n_y,n_z)$)

$$\epsilon_{\vec{n}} = \pi^2(n_x^2 + (\frac{n_y L_x}{L_y})^2 + (\frac{n_z L_x}{L_y})^2). \quad (6)$$

It is easy to take into account spin effects and dephasing by magnetic field. To describe those, we introduce the following dimensionless parameters: $\eta_{SO} = D\tau_{SO}/L_x^2$, $\tau_{SO}$ being characteristic time of spin-flip process, and $\eta = eHL_xL_y/c\hbar$, H being magnetic field $\parallel z$. We can take different $\eta$ for different thus describing parametric correlations of Then all diffusion modes acquire an extra shift $(\eta-\eta')^2$ while the cooperon ones are shifted by $(\eta+\eta')^2$. If spin-orbit interaction is taken into account, for each branch there are three degenerate eigenvalues corresponding to spin triplet and the one corresponding to spin singlet. The triplet eigenvalues acquire the same extra shift for all branches, $\epsilon^{triplet} = \epsilon^{singlet} + \eta_{SO}$.

Due to extra matrix structure of the Green's functions in use there are two eigenvalues $\mu$ corresponding to each either cooperon or diffusion mode,

$$\mu_m^\pm = \frac{l^2}{L_x^2}(\epsilon_m - \frac{(\Phi\pm\Phi')^2}{4}), \quad (7)$$

where mean free path $l=\sqrt{D\tau}$, $m$ labeling all abovementioned modes. Substituting this to 3, we find

$$\sin\Phi\sin\Phi' <F(\Phi)F(\Phi')> = 4\frac{\partial}{\partial\Phi}\frac{\partial}{\partial\Phi'}\sum_{modes}[\ln(\epsilon_m - \frac{(\Phi+\Phi')^2}{4}) + \ln(\epsilon_m - \frac{(\Phi+\Phi')^2}{4})]. \quad (8)$$

Finally we express the answer in terms of density-density correlator. It is convenient to consider the density in new coordinate $\lambda$ related to transmissions by $T=1/\sinh^2\lambda$. Using identity $2\pi\rho(\lambda)=F(2\lambda+\pi)-F(2\lambda-\pi)$ that immediately follows from 2, we obtain

$$<\rho(\lambda,\eta)\rho(\lambda',\eta')> = \frac{1}{2\pi^2}\sum_{modes} Z(\sqrt{\epsilon_m}+\pi) + Z(\sqrt{\epsilon_m}-\pi) - 2Z(\sqrt{\epsilon_m}); \quad (9)$$

$$Z(x;\lambda,\lambda') \equiv \text{Re}(\frac{1}{(x+i(\lambda+\lambda'))^2} + \frac{1}{(x+i(\lambda-\lambda'))^2}).$$

This relation provides an exact microscopic answer for transmission correlations and is the main result of the present work. It describes the effects of dimensionality and crossovers between different statistical ensembles.

Below we analyze the physically meaningful consequences of 10.

First let us consider correlations at small separations between $\lambda$'s, $|\lambda'-\lambda|\ll 1$. Them the sum is dominated by the second term corresponding $n_x=1, n_y=n_z=0$. Manipulating $\eta,\eta_{S0}$ allows to consider limits of different pure statistical ensembles and to reconcile that

$$<\rho(\lambda)\rho(\lambda')> = -\frac{s^2}{\pi^2\beta}\text{Re}\frac{1}{(\lambda-\lambda'+i0)^2}, \quad (10)$$

$s$ being an extra degeneracy of transmission eigenvalues not lifted by the random scattering. This coincides with a known result of Dyson [15] provided $|\lambda - \lambda'|$ exceeds the average spacing between transmissions, $sG_Q/G$, and thus proves Imry conjecture.

The present theory is not able to give a detailed form of the correlator at separations comparable with the average spacing, albeit we believe it coincides with Dyson expressions. This is a common drawback of theories perturbative in $G/G_q$, like, for instance, [10]. At small separations, the fluctuations of the mode $n_x = 1, n_{y,z} = 0$ become big, like the fluctuations of the mode $n_{x,y,z}$ for an isolated system. [10]. For an isolated system, the answer has been gotten by non-perturbative supersymmetry technique. [11] The development of supersymmetry technique for diagrammatic approach [9] that has been started [16] recently, will do the same for transmission.

If $\eta \neq \eta' \ll \pi$ the contribution of divergent mode

$$<\rho(\lambda,\eta)\rho(\lambda',\eta')> = -\frac{s^2}{2\pi^2}\text{Re}\frac{1}{(\lambda - \lambda' + i(\eta - \eta')/2\pi)^2}, \quad (11)$$

reconciles universal parametric correlations [17] that correspond to Brownian motion of transmissions. [18]

Thus we have shown that the transmission statistics at small separations obeys that one of Wigner-Dyson ensemble. The rest of terms in 10, however, do not have universal form, that explains why the fluctuations of the quantities integrated over transmission distribution, like conductance, do not obey universal relations.

The question of interest is if one can improve RMT in such a way that it accounts for details of the sample geometry and describes crossover between ensembles. The answer seems to be positive.

The proven way of such an improvement is so-called DMKP method [19] that seem to work for quasi-one-dimensional systems. [7] Indeed, for quasi-one-dimensional case, and for pure ensembles, $\sqrt{\epsilon_n} = \pi n$, sum in 10 become telescopic and can be readily done. The result coincides with the one obtained in [6] by DMKP method,

$$<\rho(\lambda)\rho(\lambda')> = \frac{s^2}{\pi^2\beta}\text{Re}[\frac{1}{(\lambda - \lambda' + i\pi)^2} - \frac{1}{(\lambda - \lambda' + i0)^2}], \quad (12)$$

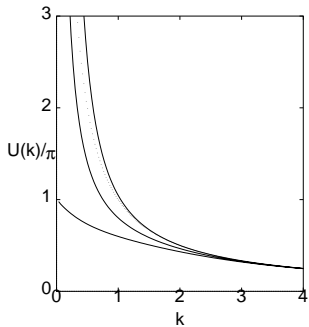

FIG. 1. The Fourier component of effective eigenvalue interaction for ( from the uppermost to lowermost curve ) wire, Wigner-Dyson interaction, square, and cube.

To become more optimistic, let us note that the correlator 10 possesses the following property,

$$<\rho(\lambda)\rho(\lambda')> = K(\lambda - \lambda') + K(\lambda + \lambda'). \quad (13)$$

In the Green's function framework, it is impossible to find a physical reason for this. RMT gives a bright pattern to comprehend such a symmetry: consider a system of classical particles confined in the region $0 < \lambda$. The correlator of such form emerges if one assumes pair interaction $U(\lambda - \lambda')$ between the particles and their "image charges" mirroring the particles at negative $\lambda$. [6] There is a simple relation between Fourier components of the interaction and $K$:

$$U(k) = \frac{s^2}{\beta K(k)} \quad (14)$$

In Fig.1 I have plotted the Fourier component of the interaction for a pure ensemble and for different geometries (wire, $L_x \gg L_{z,y}$; square, $L_x = L_y \gg L_z$; and cube, $L_x = L_y = L_z$) along with Wigner-Dyson interaction. At moderately large $k$ (small separations) for all geometries interaction rapidly converges to Wigner-Dyson values, whereas its small $k$ behavior drastically differs. For a wire, $U(k) \to k^{-2}$, and the eigenvalue repulsion exceeds the Wigner-Dyson one. For 2D geometry, $U(k) \to 2L_x/L_y|k|$. For 3D case, $U(k) \to \pi L_x^2/L_y L_z + c|k|$. This corresponds to the following asymptotes at big $\lambda$:

$$U(\lambda) \to -|\lambda| \quad \text{for 1D} \tag{15}$$

$$U(\lambda) \to -2L_x/\pi L_y \ln|\lambda| \quad \text{for 2D} \tag{16}$$

$$U(\lambda) \to 2(L_x/L_y + L_x/L_z - 1)/\lambda^2 \quad \text{for 3D.} \tag{17}$$

$$\tag{18}$$

It is interesting to note that in 3D the asymptotical interaction can be attractive for sufficiently wide slabs.

In conclusion, we have microscopically checked Imry conjecture for transmission matrix explicitly showing where and why Wigner-Dyson approach does not work. We have calculated the interaction that shall be postulated in an imaginable RMT that describes realistic diffusive conductors.

I am indebted to C. W. J. Beenakker for his persistent interest to this work, teaching me RMT and many instructive suggestions made. I appreciate interesting discussions of the results with G. E. W. Bauer, K. Frahm, and B. Rejaei. This work is a part of the research program of the "Stichting voor Fundamenteel Onderzoek der Materie" (FOM), and I acknowledge the financial support from the "Nederlandse Organisatie voor Wetenschappelijk Onderzoek" (NWO).